\documentclass[12pt]{article}
\usepackage{amssymb,latexsym}
\usepackage[dvips]{graphicx}
\begin{document}

\begin{center}
{\Large \bf
Solutions of the compatibility conditions \\[2mm]
for a Wigner quantum oscillator}\\
[5mm]
{\bf N.I.~Stoilova}\footnote{E-mail: Neli.Stoilova@UGent.be; Permanent address:
Institute for Nuclear Research and Nuclear Energy, Boul.\ Tsarigradsko Chaussee 72,
1784 Sofia, Bulgaria} {\bf and J.\ Van der Jeugt}\footnote{E-mail:
Joris.VanderJeugt@UGent.be}\\[1mm]
Department of Applied Mathematics and Computer Science,
University of Ghent,\\
Krijgslaan 281-S9, B-9000 Gent, Belgium.
\end{center}

\vskip 2cm
\noindent
Short title: Wigner quantum oscillators

\vskip 5mm
\noindent
PACS numbers: 03.65.-w,03.65.Fd,02.20.-a

\vskip 1cm
\begin{abstract}
We consider the compatibility conditions for a $N$-particle $D$-dimensional 
Wigner quantum oscillator. These conditions can be rewritten as certain
triple relations involving anticommutators, so it is natural to look
for solutions in terms of Lie superalgebras.
In the recent classification of ``generalized quantum statistics'' for
the basic classical Lie superalgebras~\cite{NJ}, each such statistics
is characterized by a set of creation and annihilation operators plus a
set of triple relations. In the present letter, we investigate which cases
of this classification also lead to solutions of the compatibility conditions.
Our analysis yields some known solutions and several classes of new solutions.
\end{abstract}

\vfill\eject

In a previous paper~\cite{NJ} we made a classification of all 
generalized quantum statistics (GQS) associated with the basic 
classical Lie superalgebras $A(m|n)$, $B(m|n)$, $C(n)$ and $D(m|n)$. 
Each such statistics is determined by $M$ creation operators $x_i^+$ ($i=1,\ldots,M$) and
$M$ annihilation operators $x_i^-$ ($i=1,\ldots,M$), which generate the corresponding  
superalgebra $G$ subject to certain triple relations ${\cal R}$. This leads to
a $\mathbb{Z}$-grading of $G$ of the form
\begin{equation}
G=G_{-2} \oplus G_{-1} \oplus G_0 \oplus G_{+1} \oplus G_{+2}, 
\label{5grading}
\end{equation}
with $G_{\pm 1}= \hbox{span}\{x^\pm_i,\ i=1,\ldots,M\}$
and $G_{j+k}=[\![ G_j,G_k ]\!]$, where $[\![\cdot,\cdot]\!]$ is the Lie superalgebra bracket.
The known cases, namely para-Bose and para-Fermi statistics~\cite{Green}, and 
$A$-(super)statistics~\cite{Palev1}-\cite{sl(1|n)} appear as simple 
examples in the classification.

In the present letter we are dealing with a different problem, namely
finding solutions of the compatibility conditions (CCs)
of a Wigner quantum oscillator system. These compatibility conditions
take the form of certain triple relations for operators.
So formally the CCs appear as special triple relations among
operators which resemble the creation and annihilation operators of
a generalized quantum statistics. One can thus investigate which
formal GQSs also provide solutions of the CCs.
It turns out that the classification presented in~\cite{NJ} yields 
new solutions of these compatibility conditions.

The concepts of Wigner quantization~\cite{Palev2} and of a Wigner Quantum System 
(WQS)~\cite{Palev3} were introduced by Palev, inspired by~\cite{Wigner}. WQSs are noncanonical generalized
quantum systems for which Hamilton's equations are identical to the Heisenberg 
equations and for which certain additional properties, valid for any quantum system, are also
fulfilled.
For more examples of WQSs and physical aspects, see~\cite{Yang}-\cite{Blasiak}.

Let us briefly describe a WQS consisting of $N$ $D$-dimensional isotropic harmonic oscillators.
The Hamiltonian of this $N$-particle $D$-dimensional ($D=1,2,3$) harmonic oscillator 
system is given by
\begin{equation}
\hat{H}=\sum_{\alpha=1}^{N} \Big({ {\hat {\bf P}}_\alpha^2 \over 2m}
+{m\omega^2\over{2}}{{\hat {\bf R}}}_\alpha^2 \Big), \label{Ham0} 
\end{equation} 
with $m$ the mass and $\omega$ the frequency of each oscillator.
The Hamiltonian ${\hat H}$ depends on the $2DN$ variables ${\hat R}_{\alpha i}$ and ${\hat P}_{\alpha i}$, 
with $\alpha =1,\ldots,N$ and $i=1,\ldots,D$. In practice, 
the cases $D=1,2,3$ will be the most interesting, but we shall treat the 
general situation here.

In a Wigner quantum system, the operators
${\hat {\bf R}}_1,\ldots,{\hat {\bf R}}_N$ and ${\hat {\bf P}}_1,\ldots,{\hat {\bf P}}_N$ have to be defined in such a
way that Hamilton's equations
\begin{equation}
    {\dot{{\hat {\bf P}}}}_\alpha=-m\omega^2{\hat {\bf R}}_\alpha, \ \ {\dot{{\hat {\bf R}}}}_\alpha = {1\over m}{\hat {\bf P}}_\alpha
    \ ~ {\rm for} ~\ \alpha=1,2,\ldots,N,
     \label{Ham}
\end{equation}
and the Heisenberg equations
\begin{equation}
     {\dot{{\hat {\bf P}}}}_\alpha = {i\over{\hbar}}[\hat{H},{\hat {\bf P}}_\alpha], \ \
     {\dot{{\hat {\bf R}}}}_\alpha = {i\over{\hbar}}[\hat{H},{\hat {\bf R}}_\alpha]
     \ ~ {\rm for} ~ \ \alpha=1,2\ldots,N,
     \label{Heis}
\end{equation}
are identical as operator equations. These compatibility conditions (CCs) are
as follows
\begin{equation}
   [\hat{H},{\hat {\bf P}}_\alpha]=i\hbar m \omega^2{\hat {\bf R}}_\alpha ,\ \
   [\hat{H},{\hat {\bf R}}_\alpha]=-{{i\hbar}\over{m}}{\hat {\bf P}}_\alpha
    \ ~ {\rm for} ~ \ \alpha=1,2,\ldots,N.
     \label{comp}
\end{equation}
To make the connection with basic classical Lie superalgebras we write the operators
${\hat {\bf P}}_\alpha$ and ${\hat {\bf R}}_\alpha$ ($\alpha=1,2,\ldots,N$) in terms of new operators (or vice versa):
\begin{equation}
a_{\alpha j}^\pm = \sqrt{cm \omega \over 4\hbar}
 {\hat R}_{\alpha j} \pm i \mu 
  \sqrt {c\over 4m \omega \hbar} {\hat P}_{\alpha j}, \qquad \label{A}
  (\alpha=1,\ldots,N;\ j=1,\ldots,D) \nonumber
\end{equation}
where $\mu=+1$ or $-1$ and $c$ is an arbitrary positive constant (which can be
chosen as an integer).
The Hamiltonian $\hat{H}$ is then 
\begin{equation}
\hat{H} = {{\omega\hbar}\over{c}}\sum_{\alpha =1}^N \sum_{i=1}^D \{a_{\alpha i}^+,a_{\alpha i}^-\},
     \label{Halpha}
\end{equation}
with $\{\cdot,\cdot\}$ an anticommutator.
The compatibility conditions~(\ref{comp}) take the form:
\begin{equation}
\sum_{\alpha=1}^N  \sum_{i=1}^D  [ \{a_{\alpha i}^+,a_{\alpha i}^- \},a_{\beta j}^\pm]
=\mp \mu c\; a_{\beta j}^\pm , \qquad (\beta =1,\ldots,N;\ j=1,\ldots,D).
\label{comp1}
\end{equation}
In the present form, the compatibility conditions are expressed as
certain triple relations for a set of odd operators $a_{\alpha i}^\pm$.
Thus it is natural to look for solutions of~(\ref{comp1}) in the framework of
Lie superalgebras. The classification of GQSs~\cite{NJ}, also expressed
by means of certain creation and annihilation operators (CAOs) $x_i^\pm$ ($i=1,\ldots,M$) satisfying
triple relations~${\cal R}$, can thus be used to investigate solutions of~(\ref{comp1}).
In the classification list of~\cite{NJ}, we should now restrict ourselves 
to cases where all CAO's of ${\cal R}$ consist of odd elements only.
Therefore $G_{-1}$ and $G_{+1}$ are odd subspaces, 
and the grading~(\ref{5grading}) is {\em consistent} with the
$\mathbb{Z}_2$-grading of the Lie superalgebra. 
Then, after identifying the $x_i^\pm$ with the operators $a_{\alpha j}^\pm$
(eventually up to an overall constant),
it remains to verify whether~(\ref{comp1}) is satisfied. 
We shall now perform this investigation for the basic classical Lie superalgebras.
 
\vskip 5mm

For the Lie superalgebra $sl(m|n)=A(m-1|n-1)$ there are two GQSs with all CAOs odd elements~\cite{NJ}.
The first of these corresponds to a grading of length~3 (i.e.\ $G_{\pm2}=0$ in~(\ref{5grading})). 
In this case, the CAOs are given by:
\begin{equation}
x_{rk}^- = e_{k, r+m+1}, \quad x_{rk}^+ = e_{r+m+1,k}, \qquad r=1,\ldots,n;\;
k=1,\ldots,m,
\end{equation}
where $e_{ij}$ is a $(m+n)\times(m+n)$ matrix with zeros everywhere except a 
$1$ on the intersection of row $i$ and column $j$ (corresponding to the defining
$sl(m|n)$ representation).
These operators satisfy the triple relations (we write in this paper only the relations 
from~${\cal R}$ that are needed here;
$r,s,t=1,\ldots,n$; $i,j,k=1,\ldots,m$)
\begin{eqnarray}
&& [\{x_{ri}^+,x_{sj}^-\},x_{tk}^+]=
\delta_{ij}\delta_{st} x_{rk}^+ -\delta_{jk}\delta_{rs} x_{ti}^+,  \nonumber\\
&& [\{x_{ri}^+,x_{sj}^-\},x_{tk}^-]=
-\delta_{ij}\delta_{rt} x_{sk}^- +\delta_{ik}\delta_{rs} x_{tj}^-, \label{sl}
\end{eqnarray}
and thus
\begin{equation}
\sum_{r=1}^n\sum_{k=1}^m  [ \{x_{rk}^+,x_{rk}^- \},x_{sj}^\pm]
=\pm (m-n) x_{sj}^\pm. \label{slmnsolution}
\end{equation}
It is clear that such systems provide solutions for the CCs (as long as $m\ne n$).
First of all, taking $m=D$ and $n=N$ yields the $sl(D|N)$ solution
of the CCs~(\ref{comp1}) for the $N$-particle $D$-dimensional oscillator,
by taking $a_{\alpha j}^\pm = x_{\alpha j}^\pm$ ($\alpha=1,\ldots,N$; $j=1,\ldots,D$).
This is (at least for $D=3$) a known solution: see~\cite{PS1} for
a discussion and some properties corresponding to this $sl(3|N)$ case.

Secondly, one can take $m=1$ and $n=DN$, yielding the $sl(1|DN)$
solution of the CCs. In this case, one takes 
$a_{\alpha j}^\pm = x_{j+(\alpha-1)D,1}^\pm$ ($\alpha=1,\ldots,N$; $j=1,\ldots,D$).
This is again a known solution: see~\cite{Palev2}, \cite{PS}-\cite{K2} for an investigation
of the physical properties of the $sl(1|3N)$ solution of the 
Wigner quantum oscillator.

Observe that one can always interchange the operators $a_{\alpha j}^+$ with $a_{\alpha j}^-$.

Note that the cases $(m=N,n=D)$ or $(m=DN,n=1)$ also provide solutions, but 
these are not considered because of the isomorphism of $sl(m|n)$ and $sl(n|m)$.
More generally, it is clear that by repartitioning the $mn$ operators $x_{rk}^+$
($r=1,\ldots,n$; $k=1,\ldots,m$) into $N$ sets of $D$ operators (and analogously
for the $x_{rk}^-$), (\ref{slmnsolution}) still yields a solution of~(\ref{comp1}).
This means that all Lie superalgebras $sl(m|n)$ with $mn=DN$ provide a solution
to the compatibility conditions for the $N$-particle $D$-dimensional Wigner quantum
oscillator.

The second type of GQS for the Lie superalgebra $sl(m|n)$ with all CAOs odd elements
corresponds to a grading of length~5~\cite{NJ}. In this situation there are several
inequivalent GQSs, all of them leading to solutions of the CCs. Since the description 
is somewhat more complicated than the other cases, we shall give it in the Appendix.

\vskip 5mm

Next, we turn our attention to the Lie superalgebras $B(m|n)=osp(2m+1|2n)$. 
We know from~\cite{NJ} that there is one GQS with odd elements only. 
In terms of the defining $(2m+2n+1)$-dimensional representation of $B(m|n)$,
the corresponding CAOs are given by:
\begin{eqnarray}
x_{ri}^+ = e_{m+i, 2m+1+r}-e_{2m+1+n+r,i}, && 
x_{ri}^- = e_{i, 2m+1+n+r}+e_{2m+1+r,m+i},\nonumber \\
x_{r, -i}^+ = e_{i, 2m+1+r}-e_{2m+1+n+r,i+m}, && 
x_{r, -i}^- = e_{m+i, 2m+1+n+r}+e_{2m+1+r,i},\nonumber \\
x_{r 0}^+ = e_{2m+1, 2m+1+r}-e_{2m+1+n+r,2m+1}, &&
x_{r 0}^- = e_{2m+1, 2m+1+n+r}+e_{2m+1+r,2m+1},\nonumber 
\end{eqnarray}
with $r=1,\ldots,n$ and $i=1,\ldots,m$. If we introduce the notation
\begin{equation}
\langle j\rangle = \left\{ \begin{array}{lll}
 {\;\; 1} & \hbox{if} & j=1,\ldots ,m \\ 
 {-1} & \hbox{if} & j=-m,\ldots ,-1 \nonumber \\
 {\;\; 0}  & \hbox{if} & j=0 \\
 \end{array}\right. 
\end{equation}
the triple relations needed can be written as follows:
\begin{equation}
[\{x_{rk}^+,x_{rk}^-\},x_{sj}^{\pm}]=
\pm \langle k\rangle \langle j\rangle \delta_{|k||j|}\; x_{sj}^{\pm}
\mp \delta_{rs}\; x_{sj}^{\pm},
\qquad(r,s =1,\ldots,n;\; k,j=-m,\ldots,m).
\end{equation}
This implies
\begin{equation}
\sum_{r =1}^{n} \sum_{k=-m}^{m}[\{x_{rk}^+,x_{rk}^-\},x_{sj}^{\pm}]=
\mp (2m+1)x_{sj}^{\pm}, \qquad (s=1,\ldots,n;\; j=-m,\ldots,m).
\label{osp-solution}
\end{equation}
Again it is clear that this provides solutions for the CCs. 
For $D=2m+1$ and $N=n$, one obtains the $osp(D|2N)$ solution of the
CCs~(\ref{comp1}) for the $N$-particle $D$-dimensional oscillator,
by taking $a^\pm_{\alpha j} = x^\pm_{\alpha j}$ ($\alpha=1,\ldots,N$; $j=-m,\ldots,m$).
This is a new class of solutions of WQSs. Note that even the 
simplest case ($D=3$ and $N=1$, or $osp(3|2)$) is different from the
$osp(3|2)$ solution of~\cite{osp32}, since in the current case the
operators $a^\pm_{\alpha j}$ correspond to root vectors of $osp(3|2)$ 
(which was not the case in~\cite{osp32}).

Alternatively, one can also take $N=2m+1$ and $D=n$ in~(\ref{osp-solution}).
This yields the $osp(N|2D)$ solution of the CCs for the $N$-particle
$D$-dimensional oscillator, by taking 
$a^\pm_{\alpha j} = x^\pm_{j \alpha}$ ($\alpha=-m,\ldots,m$; $j=1,\ldots,D$).
More generally, it is clear that by repartitioning the $(2m+1)n$ operators $x_{rk}^+$
($r=1,\ldots,n$; $k=-m,\ldots,m$) into $N$ sets of $D$ operators (and analogously
for the $x_{rk}^-$), (\ref{osp-solution}) still yields a solution of~(\ref{comp1}).
This means that all Lie superalgebras $osp(2m+1|2n)$ with $(2m+1)n=DN$ provide a solution
to the compatibility conditions.

Finally, one can have $m=0$ and $n=DN$, yielding the $B(0|DN)=osp(1|2DN)$ solution
of the CCs. In this case, one obtains a solution for the $N$-particle $D$-dimensional
oscillator, by taking $a^\pm_{\alpha j} = x^\pm_{j+(\alpha-1)D,0}$ ($\alpha=1,\ldots,N$; $j=1,\ldots,D$).
This solution is not new; in fact it is (up to a constant) the known para-Bose 
solution~\cite{Palev2}, \cite{GP}.
Indeed, let us put
\begin{equation}
b^+_r= \sqrt{2}\, x^+_{r0}, \qquad b^-_r= -\sqrt{2}\, x^-_{r0},
\end{equation}
for $r=1,\ldots,DN$. Then these operators satisfy
\begin{eqnarray}
&& [\{ b_r^{\xi}, b_s^{\eta}\} , b_t^{\epsilon}]=
(\epsilon -\xi) \delta_{rt} b_s^{\eta} +  (\epsilon -\eta)\delta_{st}b_r^{\xi},  \label{pBose} \\
&& \qquad\qquad \xi, \eta, \epsilon =\pm\hbox{ or }\pm 1;\quad r,s,t=1,\ldots,DN. \nonumber 
\end{eqnarray}
These are the para-Bose operators of~\cite{Green}. 
For $osp(1|6N)$, it was observed in~\cite{PS1} that this yields a solution of the CCs
for the $N$-particle 3-dimensional Wigner quantum oscillator. 

\vskip 5mm
Let us now consider the Lie superalgebras $D(m|n)=osp(2m|2n)$. From~\cite{NJ}
it follows that there are two GQSs with odd elements only. In terms of the
defining $(2m+2n)$-dimensional representation of $D(m|n)$, the CAOs of the first
system are given by:
\begin{eqnarray}
x_{ri}^+ = e_{m+i, 2m+r}-e_{2m+n+r,i}, &&
x_{ri}^- = e_{i, 2m+n+r}+e_{2m+r,m+i},\nonumber \\
x_{r, -i}^+ = e_{i, 2m+r}-e_{2m+n+r,i+m}, &&
x_{r, -i}^- = e_{m+i, 2m+n+r}+e_{2m+r,i}\label{D1} ,
\end{eqnarray}
with $r=1,\ldots,n$ and $i=1,\ldots,m$. It is easy to verify that these
satisfy
\begin{equation}
[\{x_{rk}^+,x_{rk}^-\},x_{sj}^{\pm}]=
\pm \langle k\rangle \langle j\rangle \delta_{|k||j|}\; x_{sj}^{\pm}
\mp \delta_{rs}\; x_{sj}^{\pm},
\qquad(r,s =1,\ldots,n;\; k,j=\pm 1,\ldots,\pm m).
\end{equation}
Thus it follows that
\begin{equation}
\sum_{r =1}^{n} \sum_{0\ne k=-m}^{m}[\{x_{rk}^+,x_{rk}^-\},x_{sj}^{\pm}]=
\mp 2m\; x_{sj}^{\pm}, \qquad (s=1,\ldots,n;\; j=\pm 1,\ldots,\pm m).
\label{osp2-solution}
\end{equation}
For $D=2m$ and $N=n$, this yields the $osp(D|2N)$ solution of the CCs~(\ref{comp1})
for the $N$-particle $D$-dimensional oscillator, by taking
$a^\pm_{\alpha j} = x^\pm_{\alpha j}$ ($\alpha=1,\ldots,N$; $j=\pm 1,\ldots,\pm m$).
This is a new class of solutions for the WQSs. 

Alternatively, one can take $N=2m$ and $D=n$ in~(\ref{osp2-solution}). 
This yields the $osp(N|2D)$ solution of the CCs~(\ref{comp1}), by taking
$a^\pm_{\alpha j} = x^\pm_{j \alpha}$ ($j=1,\ldots,D$; $\alpha=\pm 1,\ldots,\pm m$).
As before, one can more generally repartition the $2mn$ operators $x_{rk}^+$
($r=1,\ldots,n$; $k=\pm 1,\ldots,\pm m$) into $N$ sets of $D$ operators (and analogously
for the $x_{rk}^-$); then (\ref{osp2-solution}) still yields a solution of~(\ref{comp1}).
This means that all Lie superalgebras $osp(2m|2n)$ with $2mn=DN$ provide a solution
to the compatibility conditions for the $N$-particle $D$-dimensional Wigner quantum
oscillator.

The Lie superalgebra $D(m|n)=osp(2m|2n)$ also admits a different GQS with odd elements only~\cite{NJ}.
The CAOs of this second system are given by:
\begin{eqnarray}
x_{ri}^+ = e_{2m+n+i,r}-e_{m+r,2m+i}, &&
x_{ri}^- = e_{r,2m+n+i}+e_{2m+i,m+r},\nonumber \\
x_{r, -i}^+ = e_{m+r,2m+n+i}+e_{2m+i,r}, &&
x_{r, -i}^- = e_{r,2m+i}-e_{2m+n+i,m+r}\label{D2} ,
\end{eqnarray}
with $r=1,\ldots,m$ and $i=1,\ldots,n$. 
Although this looks similar to the first system, observe that it is essentially different.
In~(\ref{D1}), the subalgebra $G_0=[\![ G_{-1}, G_{+1} ]\!]$ is $sl(n)\oplus so(2m)$, 
whereas in~(\ref{D2}), it is $sl(m)\oplus sp(2n)$~\cite{NJ}.
In this case the operators~(\ref{D2}) satisfy
\begin{equation}
[\{x_{rk}^+,x_{rk}^-\},x_{sj}^{\pm}]=
\pm \langle k\rangle \langle j\rangle \delta_{|k||j|}\; x_{sj}^{\pm}
\mp \delta_{rs}\; x_{sj}^{\pm},
\qquad(r,s =1,\ldots,m;\; k,j=\pm 1,\ldots,\pm n).
\end{equation}
Now we have
\begin{equation}
\sum_{r =1}^{m} \sum_{0\ne k=-n}^{n}[\{x_{rk}^+,x_{rk}^-\},x_{sj}^{\pm}]=
\mp 2n\; x_{sj}^{\pm}, \qquad (s=1,\ldots,m;\; j=\pm 1,\ldots,\pm n).
\label{osp3-solution}
\end{equation}
For $N=2m$ and $D=n$, this yields the second $osp(N|2D)$ solution of the CCs~(\ref{comp1}), by taking
$a^\pm_{\alpha j} = x^\pm_{\alpha j}$ ($j=1,\ldots,D$; $\alpha=\pm 1,\ldots,\pm m$).
As for the other cases, one can 
more generally repartition the $2mn$ operators $x_{rk}^+$
($r=1,\ldots,m$; $k=\pm 1,\ldots,\pm n$) into $N$ sets of $D$ operators (and analogously
for the $x_{rk}^-$) and still obtain a solution of~(\ref{comp1}).
Hence all Lie superalgebras $osp(2m|2n)$ with $2mn=DN$ provide a second type of solution
to the compatibility conditions for the $N$-particle $D$-dimensional Wigner quantum
oscillator.

The solutions presented here for $D(m|n)$ remain valid also when $m=1$. In that
case, the Lie superalgebra is usually denoted by $C(n+1)$: $C(n+1)=D(1|n)=osp(2|2n)$.
In particular, $C(N+1)$ yields solutions for the $N$-particle 2-dimensional Wigner 
quantum oscillator.

\vskip 5mm
To conclude, our analysis of the compatibility conditions~(\ref{comp1}) using
the formal classification of GQS in~\cite{NJ} has given rise to several
classes of new solutions for the $N$-particle $D$-dimensional Wigner quantum
oscillator.
The most interesting solutions are those with $D=1,2,3$.
For example, for $D=1$ there are solutions in terms of the Lie superalgebras
$sl(1|N)$ and $osp(1|2N)$; for $D=2$ there are solutions in terms of 
$sl(1|2N)$, $sl(2|N)$, $osp(2|2N)$ and $osp(2N|2)$;
for $D=3$ there are solutions in terms of $sl(1|3N)$, $sl(3|N)$ and $osp(3|N)$
(apart from other types of partitioning).

In order to study physical properties of the new Wigner quantum systems
(energy spectrum, position and momentum operators, etc.)
one is lead to representation theory of the corresponding Lie superalgebra.
The class of representations should be ``unitary'', in the sense that
$(a_{\alpha j}^\pm)^\dagger = a_{\alpha j}^\mp$ must hold (by the
Hermiticity of the position and momentum operators, see~(\ref{A})).
For interesting examples with intriguing physical properties, see the 
$sl(1|3)$~\cite{Palev2},\cite{K1} (or 
$sl(1|3N)$) solution for the ($N$-particle) 3-dimensional Wigner
quantum oscillator~\cite{K2}.
With the current list of new solutions obtained in this letter,
we hope to investigate the physical properties of some of these
in the future.

\bigskip
\noindent{\bf Acknowledgments}
\medskip

\noindent
N.I.\ Stoilova was supported by a project from the Fund for Scientific Research -- Flanders (Belgium).

\section*{Appendix}

We describe here the remaining GQSs for the Lie superalgebra $sl(m|n)$ with odd CAOs only.
According to~\cite{NJ}, there are two classes. For the first class, $l$ can be any index
between $1$ and $m-1$, so assume that $l$ is fixed ($1\leq l<m$).
The CAOs are then described by the root vectors of~\cite[eq.~(3.9)]{NJ}, 
but in order to deduce solutions for the CCs we need to multiply them by some overall constant.
This gives, for $k=1,\ldots,m$ and $r=1,\ldots,n$:
\begin{equation}
x^+_{rk} = \left\{
\begin{array}{lcc}
\sqrt{|2m-n-2l|}\; e_{m+r,k} & \hbox{for} & k\leq l \\
\sqrt{|n-2l|} \; e_{k,m+r} & \hbox{for} & k>l
\end{array} \right. 
\end{equation}
and
\begin{equation}
x^-_{rk} = \left\{
\begin{array}{lcc}
\sqrt{|2m-n-2l|} \; e_{k,m+r} & \hbox{for} & k\leq l \\
\epsilon \sqrt{|n-2l|} \; e_{m+r,k} & \hbox{for} & k>l
\end{array} \right. 
\end{equation}
where $\epsilon=\hbox{sgn}((n-2l)(2m-n-2l))$. Of course, we have to assume that $l$ is such
that these factors do not vanish, i.e.\ $(n-2l)(2m-n-2l)\ne 0$.
Then, one can deduce that
\begin{equation}
\sum_{r=1}^n\sum_{k=1}^m  [ \{x_{rk}^+,x_{rk}^- \},x_{sj}^\pm]
=\mp \nu\; n(m-n) x_{sj}^\pm \label{slmn2solution}
\end{equation}
where $\nu=\hbox{sgn}(2m-n-2l)$. 
Clearly, for $m\ne n$ such systems provide solutions for the CCs
for the $N$-particle $D$-dimensional oscillator whenever $mn=DN$.

For the second class, $l$ can be any index
between $1$ and $n-1$.
Now the CAOs are described by the root vectors of~\cite[eq.~(3.8)]{NJ}, 
again multiplied by some appropriate constant.
This gives, for $k=1,\ldots,m$ and $r=1,\ldots,n$:
\begin{equation}
x^+_{rk} = \left\{
\begin{array}{lcc}
\sqrt{|2n-m-2l|}\; e_{m+r,k} & \hbox{for} & r\leq l \\
\sqrt{|m-2l|} \; e_{k,m+r} & \hbox{for} & r>l
\end{array} \right. 
\end{equation}
and
\begin{equation}
x^-_{rk} = \left\{
\begin{array}{lcc}
\sqrt{|2n-m-2l|} \; e_{k,m+r} & \hbox{for} & r\leq l \\
\epsilon \sqrt{|m-2l|} \; e_{m+r,k} & \hbox{for} & r>l
\end{array} \right. 
\end{equation}
where $\epsilon=\hbox{sgn}((m-2l)(2n-m-2l))$. Again we assume that $l$ is such
that these factors do not vanish, i.e.\ $(m-2l)(2n-m-2l)\ne 0$.
Now one can deduce that
\begin{equation}
\sum_{r=1}^n\sum_{k=1}^m  [ \{x_{rk}^+,x_{rk}^- \},x_{sj}^\pm]
=\mp \nu\; m(n-m) x_{sj}^\pm \label{slmn3solution}
\end{equation}
where $\nu=\hbox{sgn}(2n-m-2l)$. 
For $m\ne n$ such systems provide another class of solutions for the CCs
for the $N$-particle $D$-dimensional oscillator whenever $mn=DN$.

\end{document}